\documentclass{revtex4}

\usepackage{graphicx}
\usepackage{bm}
\usepackage{amssymb}
\usepackage{amsmath}
\usepackage{float}

\usepackage[english]{babel}
\usepackage[latin1]{inputenc}
\begin{document}

\title{Tests of the envelope theory for three-body forces}

\author{Lorenzo \surname{Cimino}}
\email[E-mail: ]{lorenzo.cimino@umons.ac.be}
\thanks{ORCiD: 0000-0002-6286-0722}
\author{Clara \surname{Tourbez}}
\email[E-mail: ]{clara.tourbez@student.umons.ac.be}
\author{Cyrille \surname{Chevalier}}
\email[E-mail: ]{cyrille.chevalier@umons.ac.be}
\thanks{ORCiD: 0000-0002-4509-4309}
\author{Gwendolyn \surname{Lacroix}}
\email[E-mail: ]{gwendolyn.lacroix@umons.ac.be}
\author{Claude \surname{Semay}}
\email[E-mail: ]{claude.semay@umons.ac.be}
\thanks{ORCiD: 0000-0001-6841-9850}
\affiliation{Service de Physique Nucl\'{e}aire et Subnucl\'{e}aire,
Universit\'{e} de Mons,
UMONS Research Institute for Complex Systems,
Place du Parc 20, 7000 Mons, Belgium}
\date{\today}

\begin{abstract}
Many-body forces, and specially three-body forces, are sometimes a relevant ingredient in various fields, such as atomic, nuclear or hadronic physics. As their precise structure is generally difficult to uncover or to implement, phenomenological effective forces are often used in practice. A form commonly used for a many-body variable is the square-root of the sum of two-body variables. Even in this case, the problem can be very difficult to treat numerically. But this kind of many-body forces can be handled at the same level of difficulty than two-body forces by the envelope theory. The envelope theory is a very efficient technique to compute approximate, but reliable, solutions of many-body systems, specially for identical particles. The quality of this technique is tested here for various three-body forces with non-relativistic systems composed of three identical particles. The energies, the eigenfunctions, and some observables are compared with the corresponding accurate results computed with a numerical variational method. 
\end{abstract}

\maketitle

\section{Introduction}
\label{sec:intro}

Two-body forces are the most common types of interaction considered in many-body quantum systems. But many-body forces, and specially three-body forces, are sometimes a crucial ingredient in atomic physics \cite{gatt11}, nuclear physics \cite{ishi17}, or hadronic physics \cite{ferr95,dmit01,pepi02,desp92,buis22}. The structures of many-body forces depend strongly on the system considered and they are generally difficult to uncover and difficult to implement in numerical codes. That is the reason why effective forms can be used instead to simulate at best possible these complicated many-body contributions avoiding some technical difficulties. A common structure used to this end for a $K$-body force in a $N$-body system with identical particles is given by 
\begin{equation}
\label{rK}
\sum_{\{i_1,\ldots,i_K\}}^N V \left( r_{\{i_1,\ldots,i_K\}} \right) \quad \textrm{with} \quad r_{\{i_1,\ldots,i_K\}}^2=\sum_{i < j}^{\{i_1,\ldots,i_K\}} r_{ij}^2 ,
\end{equation}
where $r_{ij}^2 =( \bm r_i-\bm r_j)^2$ and $\{i_1,\ldots,i_K\}$ is a set of $K$ particles among the $N$ possible ones, with $i_1 <\ldots< i_K$. The sum $\sum_{\{i_1,\ldots,i_K\}}^N$ runs over the $C_N^K$ different sets $\{i_1,\ldots,i_K\}$, while the sum $\sum_{i < j}^{\{i_1,\ldots,i_K\}}$ runs over the $C_K^2$ different pairs in a particular set $\{i_1,\ldots,i_K\}$, where $C_A^B$ is a usual binomial coefficient 
\begin{equation}
\label{binomial}
C_A^B = \frac{A!}{B!\,(A-B)!}.
\end{equation}
If $K=2$, the usual two-body case is recovered. The form (\ref{rK}) is often chosen in various fields \cite{gatt11,ferr95,dmit01,pepi02,buis22}. It is worth mentioning that some many-body forces cannot be covered by the form (\ref{rK}). For instance, complicated potentials can arise from meson exchanges \cite{epel08}. Or, in a tetraquark, the confinement is a real four-body interaction which cannot be reduced to the sum of six pairs two-body interactions \cite{deng14}. Let us remark that the variables $r_{\{i_1,\ldots,i_K\}}$ are proportional to $K$-body hyperradius defined in the hyperspherical harmonic expansion method \cite{fabr79,fabr83}. We will come back on this point below.

The envelope theory (ET) \cite{hall80,hall83,hall04,silv10,sema23} is a technique to compute approximate solutions, eigenvalues and eigenvectors, of many-body systems with arbitrary kinematics in $D$ dimensions. It is particularly simple to use for systems with identical particles and it can handle many-body forces of the form (\ref{rK}) \cite{sema13a,sema18}. The basic idea is to replace the Hamiltonian $H$ under study by an auxiliary Hamiltonian $\tilde H$ which is solvable. $\tilde H$ depends on auxiliary parameters which are optimised in such way that one particular eigenvalue of $\tilde H$ is as close as possible to the corresponding eigenvalue of $H$. So, a new set of auxiliary parameters must be determined for each eigenvalue of $H$. The auxiliary Hamiltonian is a many-body oscillator one, whose mass terms and coupling constants are the auxiliary parameters. It is supplemented by a constant term depending also on these auxiliary parameters. Many analytical solutions can be obtained for this kind of Hamiltonians \cite{cint21}. The main interest of this method is that the computational cost is independent from the number $N$ of particles, as it can be seen with (\ref{EAFM1}-\ref{EAFM3}). Moreover, in favourable situations, analytical upper or lower bounds can be computed. In this paper, the formalism will be developed for the following Hamiltonian for $N$ identical particles
\begin{equation}
\label{HN1K}
H=\sum_{i=1}^N T(p_i) + \sum_{\{i_1,...,i_K\}}^N V \left( r_{\{i_1,\ldots,i_K\}} \right).
\end{equation}
$T$ is the kinetic energy with $p_i = |\bm p_i|$ and $V$ is the $K$-body potential. As only the internal motion is relevant, $\sum_{i=1}^N \bm p_i = \bm 0$. Natural units, $\hbar=c=1$, are used in the following. Let us note that the Hamiltonian can contain several many-body potentials with various values of $K$. We keep here only one general many-body contribution to facilitate the presentation. For two-body forces in three dimensions, quite good approximations for energies and some observables can be obtained for various systems containing up to 10 identical bosons and for the spectra of three-body systems \cite{sema15a}. The accuracy can be improved, but to the detriment of the possible variational character \cite{sema15a,sema15b,lore23}. 

The purpose of this work is to test the quality of the approximate solutions obtained with the ET for the Hamiltonians (\ref{HN1K}) for several systems with $D=N=K=3$, for which accurate numerical computations can be performed. In Sec.~\ref{sec:ete}, the general equations for the ET are presented with some analytical bounds. The procedure to improve the approximation is described in Sec.~\ref{sec:impet}. The ET approximate eigenstates are built in the case $D=N=3$ in Sec.~\ref{sec:eigenstates}. Tests for the quality of the approximations for the energies, the wavefunctions and some observables are performed in Sec.~\ref{sec:tests}. Concluding remarks are given in the last section. Some calculations about the wavefunctions and the observables are given in appendix. 
 
\section{Envelope theory equations}
\label{sec:ete}

As mentioned in the introduction, the ET approximations for the Hamiltonian (\ref{HN1K}) can be computed by an optimisation procedure on some auxiliary parameters. But it is equivalent to solve this set of three equations
\begin{align}
\label{EAFM1}
E &= N\, T(p_0) + C_N^K\, V\left(\sqrt{C_K^2}\, \rho_0\right), \\
\label{EAFM2}
\sqrt{C_N^2}\, \rho_0\, p_0 &= Q , \\
\label{EAFM3}
N\, p_0 \,T'\left(p_0 \right) &=
C_N^K\, \sqrt{C_K^2}\, \rho_0\, V' \left(\sqrt{C_K^2}\, \rho_0 \right) ,
\end{align}
by eliminating the two intermediate quantities $\rho_0$ and $p_0$. $E$ is an approximation of an eigenenergy, $A'(z)=dA(z)/dz$, and the global quantum number 
\begin{equation}
\label{Q}
Q= \sum_{i=1}^{N-1}\left( 2\, n_i+l_i+\frac{D}{2} \right)
\end{equation}
determines the particular state considered. A very detailed description of the method is given in \cite{silv10} for $K=2$. Supplementary details for $K>2$ can be found in \cite{sema18}. Let us note that the two quantum numbers $\{n_i,l_i\}$ are associated with the $i$th Jacobi coordinate $\bm x_i$ \cite{silv10}, with 
\begin{equation}
\label{jacobi}
\bm x_1 = \bm r_1 - \bm r_2, \quad 
\bm x_2 = \frac{\bm r_1+\bm r_2}{2} - \bm r_3, \quad
\ldots, \quad
\bm x_N = \frac{1}{N} \sum_{i=1}^N \bm r_i, 
\end{equation}
the last coordinate being the centre of mass position. The energies $E$ correspond to completely (anti)symmetrised states \cite{silv10} (see Sec.~\ref{sec:eigenstates}). Both $\rho_0$ and $p_0$ have a nice physical interpretation: $\rho_0^2$ is the mean square distance between two particles and $p_0^2$ is the mean square momentum of each particle in the centre of mass frame for the approximate eigenvector given by the ET and associated with $E$. In some previous papers about the ET, the variable $r_0= \sqrt{C^2_N} \, \rho_0$ is used because one-body potentials were also considered. Let us note that the allowed values for $Q$ depend on the parity, the total angular momentum and the symmetry of the eigenstates (see Sec.~\ref{sec:eigenstates}).
 
For peculiar choices of $T$ and $V$, $E$ can be a lower bound or an upper bound of the exact eigenvalue. A procedure to verify if such a favourable situation happens is to define two functions $b_T$ and $b_V$ such that 
\begin{equation}
\label{hg}
T(x) = b_T(x^2) \  \textrm{and} \  V(x) = b_V(x^2).
\end{equation}
It can be shown that $E$ is an upper bound  if $b_T''(x)$ and $b_V''(x)$ are both concave functions \cite{hall83}. Conversely, if both second derivatives are convex functions, $E$ is a lower bound. If the second derivative is vanishing for one of these functions, the variational character is solely ruled by the convexity of the other. In the other cases, the variational character of the solution cannot be guaranteed. 

If several potentials with different values of $K$ are considered, their contributions must be simply added in the r.h.s. of (\ref{EAFM1}) and (\ref{EAFM3}). The energy keeps a variational character if the second derivatives of the corresponding $b$-functions have all the same convexity.

Analytical bounds exist for some systems with a power-law kinetic energy $T(p) = F \, p^\alpha$ \cite{sema18}. In order that $T$ is definite positive and increases with the momentum, $F > 0$ and $\alpha > 0$. A first example is the power-law potential 
\begin{equation}
\label{Vxpow}
V(r) = a \,\textrm{sgn}(b)\, r^b \quad \textrm{with} \quad a > 0,\ b > -\alpha,
\end{equation}
for which 
\begin{equation}
\label{Vxpow2}
\rho_0=\left[ \frac{\alpha\, N\, F\,Q^\alpha}{a\,|b|\,C_N^K\,(C_N^2)^{\alpha/2}\,(C_K^2)^{b/2}}  \right]^{1/(b+\alpha)}, 
\end{equation}
and
\begin{equation}
\label{Expow}
E=\textrm{sgn}(b)\,(b+\alpha)\left[ \left(\frac{N\,F}{|b|}\right)^b \left(\frac{a\,C_N^K}{\alpha}\right)^\alpha \left(\frac{C_K^2}{C_N^2}\right)^{\alpha\,b/2} Q^{\alpha\,b} \right]^{1/(b+\alpha)}.
\end{equation}
These energies are upper bounds if $\alpha \le 2$ and $b \le 2$. 

Another example is a potential well with the following general exponential structure 
\begin{equation}
\label{Vexp}
V(r) = -a \,\exp\left(-b\,r^{\gamma}\right) \quad \textrm{with} \quad a, b, \gamma  > 0. 
\end{equation}
With 
\begin{equation}
\label{Vexpd}
\delta = -\frac{\gamma}{\alpha + \gamma}
        \left( \frac{\alpha\, b^{\alpha/\gamma} F }{a\, \gamma}
        \frac{N}{C_N^K} \left(\frac{C_K^2}{C_N^2} \right)^{\alpha/2}
        Q^{\alpha}\right)^{\gamma/(\alpha + \gamma)} ,
\end{equation}
$\rho_0$ is given by
\begin{equation}
\label{Vexp2}
\rho_0=\frac{1}{\sqrt{C_K^2}} \left(  -\frac{\alpha+\gamma}{\gamma \, b} W_0(\delta)\right)^{\frac{1}{\gamma}}, 
\end{equation}
where $W_0$ is a branch of the multivalued Lambert function \cite{corl96}. The approximate energies are given by 
\begin{equation}
\label{Eexp}
E = -a \, C_N^K \exp\left( 
   \frac{\alpha + \gamma}{\gamma} W_0(\delta) \right)
\left[ \frac{\alpha + \gamma}{\alpha} W_0(\delta) + 1 \right].
\end{equation}
Thanks to the properties of $W_0$, no bound state exists for too high values of the global quantum number $Q$, as expected for such a potential well. An upper bound is obtained if $\alpha \le 2$ and $\gamma \le 2$.

\section{Modified envelope theory}
\label{sec:impet}

The main drawback of the ET is the unphysical strong degeneracy of the global quantum number $Q$ inherent to the method. A procedure to improve the approximations is to replace $Q$ into the system (\ref{EAFM1}-\ref{EAFM3}) by a new global quantum number $Q_\phi$ containing a parameter $\phi$ and defined by
\begin{equation}
\label{Qphi}
Q_\phi = \sum_{i=1}^{N-1} \left(\phi\, n_i + l_i + \frac{D+\phi-2}{2} \right),
\end{equation}
which partly breaks the strong degeneracy of $Q$. This relies on the existence of universal effective quantum number for centrally symmetric two-body problems \cite{loba09}. Equation~(\ref{Qphi}) can be considered as a generalisation to $N$-body systems of this idea. The parameter $\phi$ can be determined by a fit on a single known accurate solution, for instance the ground state. This procedure can give very good improvements in some cases \cite{sema15a}. The genuine ET, with its possible variational solutions, is recovered with $\phi=2$. For other values of $\phi$, the variational character of the solution cannot be guaranteed.

Combining the ET with a generalisation to $N$-body systems of the dominantly orbital state (DOS) method \cite{olss97}, it is possible to provide a value of $\phi$ without knowing an accurate solution. This procedure gives very good results for two-body forces \cite{sema15b,chev22,lore23}. The generalisation to $K$-body forces follows the same tracks. The formula is given by 
\begin{equation}
\label{phiid}
\phi_\textrm{DOS} = \sqrt{\frac{k}{N\,C_N^2\, {\tilde p_0}^3\, T'(\tilde p_0)}}Q_0,
\end{equation}
with $k$ and $Q_0$ given by 
\begin{align}
\label{DeltaE}
k &= \frac{2\, N\,\tilde p_0}{\tilde \rho_0^2}\,T'(\tilde p_0)
+ \frac{N\,\tilde p_0^2}{\tilde \rho_0^2}\,T''(\tilde p_0)
+  C_K^2\,C^K_N\, V''\left(\sqrt{C_K^2}\,\tilde \rho_0\right), \\
Q_0 &= \sum_{i=1}^{N-1}\left (l_i + \frac{D-2}{2} \right).
\end{align}
The parameters $\tilde \rho_0$ and $\tilde p_0$ are solutions of the system (\ref{EAFM1}-\ref{EAFM3}) but with $Q$ replaced by $Q_0$ which determines the excitation due to the angular momentum only. A particular state is now fixed by the choices of $Q$ and $Q_0$. When $K=2$, the formulas of \cite{chev22} are recovered. In some cases, the value of $\phi_\textrm{DOS}$ can be very simple. For instance, 
\begin{equation}
\label{phispec}
\phi_\textrm{DOS} = \sqrt{\alpha+b}
\end{equation}
for the energy given by (\ref{Expow}). This implies that $\phi_\textrm{DOS}=2$ in the case of a many-body harmonic oscillator ($N$ and $K$ arbitrary) and $\phi_\textrm{DOS}=1$ for the Kepler problem ($N=K=2$), both values giving the exact energy in these two particular cases. For the energy given by (\ref{Eexp}), the form of $\phi_\textrm{DOS}$ is quite complicated and it is computed numerically.

Even with the use of $Q_\phi$ instead of $Q$ to compute the energies, some degeneracies remain. It is thus not immediate to identify spectra computed with the ET with numerical accurate spectra or with experimental spectra where these non physical degeneracies are missing. If the ET energies are lower or upper bounds, this variational character can be used to identify the different levels. The parity and the possible values of the total angular momentum (see next section) compatible with the quantum numbers $\{ l_i \}$ can also help. 

\section{Envelope theory eigenstates for $\bm{D=3}$}
\label{sec:eigenstates}

\subsection{Many-body eigenstates}
\label{sec:mbe}

To check the quality of the ET approximations it is interesting to also look at the wavefunctions provided by the method. For $N=2$, the comparison is easy, and good results for various potentials are presented in \cite{sema10}. For $N>2$, it is more convenient to look at some observables as in \cite{sema15a,sema15b}. In this work, we only consider space degrees of freedom for $D=3$. Then, an ET approximate $N$-body eigenstate is built from the solutions of the auxiliary Hamiltonian which is a $N$-body harmonic oscillator Hamiltonian \cite{silv12}
\begin{equation}\label{psiphi}
\Psi=\phi_{\textrm{cm}}(\bm x_N)\,\prod^{N-1}_{i=1}\, \varphi_{n_i l_i m_i}(\bm y_i),
\end{equation}
where $\varphi_{n_i l_i m_i}(\bm y_i)$ is a 3-dimensional harmonic oscillator wavefunction, depending on the dimensionless variables $\bm y_i = \lambda_i\, \bm x_i$, where the variables $\bm x_i$ are the Jacobi coordinates~(\ref{jacobi}). The function $\phi_{\textrm{cm}}(\bm x_N)$ which describes the centre of mass motion can be ignored for the computation of observables. The scale parameters $\lambda_i$ are given by
\begin{equation}\label{laj}
\lambda_i=\sqrt{\frac{i}{i+1}\frac{2}{N-1} Q}\,\frac{1}{\rho_0}.
\end{equation}
A state~(\ref{psiphi}) has neither a defined total angular momentum nor a good symmetry, but it is characterised by a parity $P=(-1)^{\sum_{i=1}^{N-1} l_i}=(-1)^{Q^*}$ with $Q^*=Q-3(N-1)/2$ (this number $Q^*$ is sometimes called the excitation band). By combining such states with the same value of $Q$ (and so the same parity), it is possible, but not guaranteed, to build a physical state with a good parity, a good total angular momentum and good symmetry properties. The total angular momentum $L$ is such that $L_\textrm{min} \le L \le L_\textrm{max}=Q^*$ with $L_\textrm{min}=0$ ($1$) if $Q^*$ is even (odd). Precautions must be taken if the modified version of the ET is used for the computation of the eigenfunctions and the observables. This is explained in the next section. 

\subsection{Three-body eigenstates}
\label{sec:3be}

Let us consider first the original version of the ET. For $N=3$, a symmetrised internal wavefunction with a good parity and a good total angular momentum $L$ is written 
\begin{equation}\label{psiN3}
\psi=\sum_s c_s(Q^*) \,\psi_s \quad \textrm{with} \quad \psi_s = \left[\varphi_{n_1 l_1}(\bm y_1)\,\varphi_{n_2 l_2}(\bm y_2)\right]^L,
\end{equation}
where each $\psi_s$ is normalised and where the coefficient $c_s(Q^*)$ can be computed with the procedure described in Sec.~\ref{sec:sym}. A symbol $c_s(Q^*)$ is real and depends on the set $\{n_1,l_1,n_2,l_2,L \}$. All $c_s(Q^*)$ in the sum are characterised by the same value of $Q^*=Q-3=2 (n_1+n_2)+l_1+l_2$. The parity is $P=(-1)^{Q^*}$. The square brackets indicate the coupling, and the magnetic quantum numbers are omitted. It is worth noting that the coefficients $c_s(Q^*)$ have a purely geometrical origin, independent from the parameters of the auxiliary Hamiltonian. So, imposing $\sum_s c_s(Q^*)^2 = 1$, their values are fixed by the symmetry and the values of $Q^*$ and $L$, without any freedom. For some configurations, all $c_s(Q^*)$ can be vanishing and the corresponding state does not exist. Only the size of this wavefunction depends on the particular system considered through the coefficients $\lambda_i$ which depend on $Q$ and $\rho_0$ by (\ref{laj}).

In this paper, we focus on the observables 
\begin{equation}\label{obs1}
\langle r^k \rangle = \frac{1}{C_3^2} \sum_{i<j}^3 \langle \psi| |\bm r_i - \bm r_j|^k |\psi\rangle = \langle \psi| |\bm x_1|^k |\psi\rangle.
\end{equation}
The last equality is due to the symmetry of the wavefunction and the choice of the Jacobi coordinates. Using the orthonormality of the $\varphi_{n_i l_i m_i}(\bm y_i)$ and the fact that all functions in the sum~(\ref{psiN3}) are characterised by the same value of $Q^*$, (\ref{obs1}) reduces to
\begin{equation}\label{obs2}
\langle r^k \rangle = \sum_s c_s(Q^*)^2\, \langle r^k \rangle_{n_1(s) l_1(s)},
\end{equation}
with the notation~(\ref{rk1}) and where the indication $(s)$ for $n_1$ and $l_1$ means that these quantum numbers are those attached to the state $\psi_s$ in the sum~(\ref{psiN3}). A lengthy calculation using~(\ref{rk4}) shows that 
\begin{equation}\label{r2}
\langle r^2 \rangle = \rho_0^2,
\end{equation}
in agreement with the formalism of the ET \cite{sema15b,sema18}. Making a supplementary approximation, it is then possible to write $\langle r^k \rangle \approx \rho_0^k$, and more generally 
\begin{equation}\label{fr}
\langle f(r) \rangle \approx f(\rho_0).
\end{equation}
The advantage of this procedure is that it avoids the computation of the coefficients $c_s(Q^*)$. This last task can be very heavy if $N > 3$. In the following, the compact notation $|\sigma; Q^*; L^P\rangle$ is used for a state completely symmetrised ($\sigma=1$ ($-1$) for (anti)symmetric states) in the excitation band $Q^*$ with a good total angular momentum $L$ and a parity $P=(-1)^{Q^*}$. With the same idea, the notation $|n_1,l_1,n_2,l_2;L\rangle$ is used for a state $\psi_s$. So, (\ref{psiN3}) is written
\begin{equation}
\label{sym4}
|\sigma; Q^*; L^P\rangle = \sum_s c_s(Q^*) |n_1,l_1,n_2,l_2;L\rangle .
\end{equation}

Within the modified version of the ET, a particular state is fixed by the choices of $Q$ and $Q_0=l_1+l_2+1$ which allow the computation of $\phi$, $Q_\phi=\phi(n_1+n_2+1)+Q_0$ and the corresponding value of $\rho_0$, and finally $E$. Observables can then be obtained with (\ref{fr}). Problems appear for the calculation of eigenfunctions or observables with (\ref{obs2}). An example is given in Sec.~\ref{sec:sym} with the state $|1;3;1^-\rangle$ which is a superposition of states $\{ |0,0,1,1;1\rangle,|1,0,0,1;1\rangle\}$ with $Q_0=1$ and $|0,2,0,1;1\rangle$ with $Q_0=3$. How to fix the value of $\phi$ for such states characterised by different values of $Q_0$? A procedure can be built by pondering different contributions in the wavefunction computed with different values of $\phi$ for each value of $Q_0$. But this appear quite artificial and different trials did not bring real improvements. So, for this kind of states, it seems irrelevant to compute observables with (\ref{obs2}).

\section{Tests for several systems}
\label{sec:tests} 

In this section, the ET is tested with several three-body potentials within the framework of a nonrelativistic kinematics for $D=N=3$. So, the unique many-body variable is 
\begin{equation}\label{r123}
r_{\{123\}} = \sqrt{(\bm r_1-\bm r_2)^2 + (\bm r_1-\bm r_3)^2 + (\bm r_2-\bm r_3)^2}.
\end{equation}
It is always possible to scale the length and the energy to build dimensionless Hamiltonians. This allows to use arbitrary units, with  $T(p)=p^2/2$ for the kinetic parts with a unit mass, and allows to choose arbitrary values for some parameters of the potential (and thus for the only parameter of a power-law potential). So, the general form of the Hamiltonians studied below is
\begin{equation}\label{H123}
H=\frac{1}{2}\sum_{i=1}^3 \bm p_i^2 + V\left( r_{\{123\}} \right),
\end{equation}
where all variables $\bm p_i$, $\bm r_i$, and $r_{\{123\}}$ are dimensionless, as well as the parameters of the potential.

The results coming from the ET are compared with the accurate numerical results computed with a variational method based on an expansion with harmonic oscillator bases \cite{nunb77,silv96,silv00}. Its extension for three-body forces is described in \cite{chev23}. In this method, a trial state is given by
\begin{equation}\label{psinum}
\psi_\textrm{trial}=\sum_{Q^*}\, \sum_s d_s(Q^*) \,\left[\varphi_{n_1 l_1}(z\,\bm x_1)\,\varphi_{n_2 l_2}(2z\,\bm x_2/\sqrt{3})\right]^L,
\end{equation}
where the sum on $Q^*$ runs by step of 2 from 0 (1) for positive (negative) parity states to $Q^*_\textrm{max}$. The coefficients $d_s(Q^*)$ are determined by diagonalising the Hamiltonian matrix, and the nonlinear variational parameter $z$ is fixed by a minimisation of the energy. The coefficient $2/\sqrt{3}$ in (\ref{psinum}) is necessary to insure a good symmetry. This numerical method has been chosen because the form~(\ref{psinum}) can be directly compared with the form~(\ref{psiN3}). If the ET wavefunction is a good approximation of the exact wavefunction, one can expect that $d_s(Q^*) \approx c_s(Q^*)$ with $\sum_s d_s(Q^*)^2 \approx 1$ for the value of $Q^*$ associated with (\ref{psiN3}) and $d_s(Q^*) \ll 1$ for other values of $Q^*$. As $\bm\rho_1 = \lambda_1\, \bm x_1$ and $\bm\rho_2 = 2\lambda_1\, \bm x_2/\sqrt{3}$, we can define the parameter 
\begin{equation}\label{nu}
\nu = \frac{\lambda_1}{z} = \sqrt{\frac{Q}{2}}\frac{1}{\rho_0\, z}
\end{equation}
to measure the wavefunction size difference between (\ref{psinum}) and (\ref{psiN3}). If these two forms are close, we can expect that $\nu\approx 1$. It is checked that a value of $Q^*_\textrm{max}=18$ insures a relative accuracy around $10^{-3}$ for the linear and the Gaussian potentials. It is necessary to push to $Q^*_\textrm{max}=30$ for the Coulomb potential.
 
Besides energies $E$ and eigenstates, observables $\langle r^{-1} \rangle$, $\langle r \rangle$ and $\langle r^2 \rangle$ are computed, with (\ref{obs2}) and its approximation~(\ref{fr}). Some quantities are computed for three values of the parameter $\phi$ of Sec.~\ref{sec:impet}: $\phi=2$ which is the value given by the general theory, $\phi_\textrm{DOS}$ which is the value predicted by (\ref{phiid}) and $\phi_\textrm{GS}$ which is the value giving the exact energy for the ground state. In order to lighten the presentation, we only present results for the lowest completely symmetric states. Similar results are obtained for the antisymmetric ones. For the same purpose, some accurate results computed with the variational method are omitted, but the relative errors on the ET approximations are mentioned. The quantities $\nu$ and $\sum_s d_s(Q^*)^2$ are only shown for the computation with $\phi=2$.

As in this work we test only three identical nonrelativistic particles with only three-body forces, the interaction in Hamiltonian~(\ref{H123}) depends only on the three-body hyperradius $r_{\{123\}}$ and the solutions can be computed within the hyperspherical harmonic expansion method. It worth mentioning that very good results can be obtained by taking into account only the hyperspherical harmonics of minimal order \cite{fabr79,fabr83}.

\subsection{Linear three-body potential}
\label{sec:linear} 

The first three-body potential tested is the linear potential 
\begin{equation}\label{lin1}
V_\textrm{L}\left( r_{\{123\}} \right) = \frac{1}{2}\, r_{\{123\}} .
\end{equation}
The factor $1/2$ is arbitrary. According to (\ref{phispec}), $\phi_\textrm{DOS}=\sqrt{3}$ as $\alpha=2$ and $b=1$. A potential proportional to $r^2_{\{123\}}$ with appropriate energy scale and colour factor is a good candidate for a confining three-body interaction in a baryon \cite{ferr95,dmit01,pepi02}, but an interaction proportional to $r_{\{123\}}$ seems more realistic \cite{buis22,silv96}. Results for states with $Q^* \le 3$ are given in Tables~\ref{tab:lin1} to \ref{tab:lin5}. Several comments can be made:

\begin{itemize}
  \item Five bound states are found for $Q^* \le 3$ which correspond to the five first bound states computed numerically, with the good hierarchy. 
  \item The original method ($\phi=2$) gives good results for the energies and the observables with quite small relative errors. All energies computed with $\phi=2$ are upper bounds and the variational character is generally lost otherwise, as expected. 
  \item With $\phi=\phi_\textrm{GS}$, the energies are improved and the hierarchy respected but the observables are slightly worse. 
  \item The value of $\phi_\textrm{GS}$ shows that energies are improved with $\phi <2$. This is the case for $\phi_\textrm{DOS}$ but the value predicted is so low that it provides energies and observables which are less good than for $\phi=2$. 
  \item The accuracy of observables $\langle r^k \rangle$ is better with formula~(\ref{obs2}) than with approximation~(\ref{fr}). But the quantity $\rho_0^k$ can very easily give at least a rough approximation, and sometimes a reasonable one.
  \item For the original method ($\phi=2$), $\nu\approx 1$ and $\sum_s d_s(Q^*)^2 \approx 1$. This shows a good quality for the ET wavefunction, in agreement with the quality of energies and observables.
\end{itemize} 

\begin{table}[H]
    \begin{center}
    \begin{tabular}{cccccccc}
        \hline
         $\phi$ & $E$ & $\langle r \rangle$ & $\rho_0$ &$\langle r^2 \rangle$ & $\rho_0^2$ & $\langle r^{-1} \rangle$ & $\rho_0^{-1}$\\
        \hline
        2 &2.835[2.98]&2.011[1.21]&2.182[7.23]&4.762[3.45]&4.762[3.45]&0.633[0.02]&0.458[27.6]\\
        $\sqrt{3}$ (DOS)&2.663[3.25]&1.979[2.74]&2.050[0.74]&4.616[6.41]&4.204[14.8]&0.643[1.59]&0.488[23.0]\\
        1.871 (GS)&2.753[0.0]&1.996[1.93]&2.119[4.13]&4.693[4.85]&4.491[8.95]&0.638[0.76]&0.472[25.5]\\
        \hline
    \end{tabular} 
    \caption{Results for potential~(\ref{lin1}). Energy $E$ and three observables for the state $|1; 0; 0^+\rangle$ with $\phi=2$, $\phi_\textrm{DOS}$ and $\phi_\textrm{GS}$. Observables $\langle r^k \rangle$ are computed with (\ref{obs2}) and (\ref{fr}). Relative errors (in \%) are written between square brackets. The accurate value of the energy is $E_\textrm{acc}=2.753$. For $\phi=2$, $\nu=1.09$ and $\sum_s d_s(0)^2 = 0.97$. \label{tab:lin1}}
    \end{center}
\end{table}

\begin{table}[H]
\begin{center}
    \begin{tabular}{cccccccc}
        \hline
        $\phi$ & $E$ & $\langle r \rangle$ & $\rho_0$ &$\langle r^2 \rangle$ & $\rho_0^2$ & $\langle r^{-1} \rangle$ & $\rho_0^{-1}$\\
        \hline
        2 &3.985[5.00]&2.737[2.47]&3.068[9.33]&9.410[4.59]&9.410[4.59]&0.533[1.28]&0.326[38.1]\\
        $\sqrt{3}$ (DOS)&3.695[2.64]&2.685[4.29]&2.844[1.38]&9.061[8.13]&8.090[18.0]&0.543[3.21]&0.352[33.2]\\
        1.871 (GS)&3.847[1.35]&2.712[3.32]&2.961[5.54]&9.246[6.26]&8.769[11.1]&0.538[2.18]&0.338[35.8]\\
        \hline
    \end{tabular} 
    \caption{Same as Table~\ref{tab:lin1} for the state $|1; 2; 0^+\rangle$ with $E_\textrm{acc}=3.795$, $\nu=1.10$ and $\sum_s d_s(2)^2 = 0.94$. \label{tab:lin2}}
\end{center}
\end{table}

\begin{table}[H]
\begin{center}
    \begin{tabular}{cccccccc}
        \hline
        $\phi$ & $E$ & $\langle r \rangle$ & $\rho_0$ &$\langle r^2 \rangle$ & $\rho_0^2$ & $\langle r^{-1} \rangle$ & $\rho_0^{-1}$\\
        \hline
        2 &3.985[1.79]&2.846[0.72]&3.068[7.01]&9.410[2.12]&9.410[2.12]&0.446[0.02]&0.326[26.9]\\
        $\sqrt{3}$ (DOS)&3.841[1.88]&2.820[1.63]&2.957[3.15]&9.239[3.90]&8.744[9.05]&0.450[0.91]&0.338[24.2]\\
        1.871 (GS)&3.916[0.03]&2.834[1.15]&3.015[5.16]&9.329[2.97]&9.088[5.47]&0.448[0.42]&0.332[25.6]\\
        \hline
    \end{tabular} 
    \caption{Same as Table~\ref{tab:lin1} for the state $|1; 2; 2^+\rangle$ with $E_\textrm{acc}=3.915$, $\nu=1.06$ and $\sum_s d_s(2)^2 = 0.97$. \label{tab:lin3}}
\end{center}
\end{table}

\begin{table}[H]
\begin{center}
    \begin{tabular}{cccccccc}
        \hline
        $\phi$ & $E$ & $\langle r \rangle$ & $\rho_0$ &$\langle r^2 \rangle$ & $\rho_0^2$ & $\langle r^{-1} \rangle$ & $\rho_0^{-1}$\\
        \hline
        2 &4.500[1.49]&3.178[0.60]&3.464[8.34]&12.00[1.78]&12.00[1.78]&0.437[0.02]&0.289[34.0]\\
        $\sqrt{3}$ (DOS)&4.228[4.65]& &3.255[1.79]& &10.59[13.3]& &0.307[29.7]\\
        1.871 (GS)&4.370[1.44]& &3.364[5.21]& &11.32[7.37]& &0.297[32.0]\\
        \hline
    \end{tabular} 
    \caption{Same as Table~\ref{tab:lin1} for the state $|1; 3; 1^-\rangle$ with $E_\textrm{acc}=4.434$, $\nu=1.02$ and $\sum_s d_s(3)^2 = 0.99$. Some observables are not computed (see text). \label{tab:lin4}}
\end{center}
\end{table}

\begin{table}[H]
\begin{center}
    \begin{tabular}{cccccccc}
        \hline
        $\phi$ & $E$ & $\langle r \rangle$ & $\rho_0$ &$\langle r^2 \rangle$ & $\rho_0^2$ & $\langle r^{-1} \rangle$ & $\rho_0^{-1}$\\
        \hline
        2 &4.500[1.49]&3.272[0.60]&3.464[5.23]&12.00[1.78]&12.00[1.78]&0.367[0.02]&0.289[21.3]\\
        $\sqrt{3}$ (DOS)&4.365[1.56]&3.247[1.35]&3.360[2.07]&11.82[3.26]&11.29[7.58]&0.370[0.75]&0.298[18.9]\\
        1.871 (GS)&4.435[0.03]&3.260[0.96]&3.414[3.71]&11.91[2.49]&11.66[4.58]&0.368[0.34]&0.293[20.2]\\
        \hline
    \end{tabular} 
    \caption{Same as Table~\ref{tab:lin1} for the state $|1; 3; 3^-\rangle$ with $E_\textrm{acc}=4.434$, $\nu=1.02$ and $\sum_s d_s(3)^2 = 0.99$. \label{tab:lin5}}
\end{center}
\end{table}

\subsection{Coulomb three-body potential}
\label{sec:coulomb} 

Up to our knowledge, the following Coulomb-type interaction has no physical relevance but it can be used as a convenient approximation \cite{ferr95}. It allows here to test the method with a singular potential
\begin{equation}\label{coul1}
V_\textrm{C}\left( r_{\{123\}} \right) = -\frac{3}{r_{\{123\}}}  .
\end{equation}
The factor $3$ is arbitrary. According to (\ref{phispec}), $\phi_\textrm{DOS}=1$  as $\alpha=2$ and $b=-1$. Results for states with $Q^* \le 2$ are given in Tables~\ref{tab:coul1} to \ref{tab:coul3}. Comments about these results are:

\begin{itemize}
  \item Three bound states are found for $Q^* \le 2$ which correspond to the three first bound states computed numerically, with the good hierarchy.
  \item The original method ($\phi=2$) gives bad results for the energies and the observables, as expected from the results for two-body Coulomb potentials \cite{sema15a,lore23}. Relative errors are significantly greater than for the linear potential. All energies computed with $\phi=2$ are upper bounds and the variational character is generally lost otherwise, as expected. 
  \item With $\phi=\phi_\textrm{GS}$, the energies can be strongly improved and the hierarchy respected but the observables are sometimes better sometimes worse. 
  \item The value of $\phi_\textrm{GS}$ shows that energies are improved with $\phi <2$. This is the case for $\phi_\textrm{DOS}$ but the value predicted is so low that it provides energies and observables which are less good than for $\phi=2$ in some cases, as for the linear potential. 
  \item The accuracy of observables $\langle r^k \rangle$ is better with formula~(\ref{obs2}) than with approximation~(\ref{fr}). But the relation $\rho_0^k$ can very easily give at least a rough approximation.
  \item For the original method ($\phi=2$), $\nu$ and $\sum_s d_s(Q^*)^2$ are not so close to 1 than for the linear interaction. This is coherent with the poor quality of energies and observables.
\end{itemize} 

\begin{table}[H]
    \begin{center}
    \begin{tabular}{cccccccc}
        \hline
         $\phi$ & $E$ & $\langle r \rangle$ & $\rho_0$ &$\langle r^2 \rangle$ & $\rho_0^2$ & $\langle r^{-1} \rangle$ & $\rho_0^{-1}$\\
        \hline
        2 &$-0.167$[30.4]&4.787[15.0]&5.196[24.9]&27.00[23.6]&27.00[23.6]&0.266[19.7]&0.192[41.2]\\
        1 (DOS)&$-0.375$[56.5]&2.606[37.4]&2.309[44.5]&8.000[63.4]&5.333[75.6]&0.489[47.6]&0.433[30.8]\\
        1.502 (GS)&$-0.240$[0.0]&3.647[12.4]&3.615[13.1]&15.67[28.3]&13.07[40.2]&0.349[5.5]&0.277[16.5]\\
        \hline
    \end{tabular} 
    \caption{Results for potential~(\ref{coul1}). Energy $E$ and three observables for the state $|1; 0; 0^+\rangle$ with $\phi=2$, $\phi_\textrm{DOS}$ and $\phi_\textrm{GS}$. Observables $\langle r^k \rangle$ are computed with (\ref{obs2}) and (\ref{fr}). Relative errors (in \%) are written between square brackets. The accurate value of the energy is $E_\textrm{acc}=-0.240$. For $\phi=2$, $\nu=1.36$ and $\sum_s d_s(0)^2 = 0.90$. \label{tab:coul1}}
    \end{center}
\end{table}

\begin{table}[H]
\begin{center}
    \begin{tabular}{cccccccc}
        \hline
        $\phi$ & $E$ & $\langle r \rangle$ & $\rho_0$ &$\langle r^2 \rangle$ & $\rho_0^2$ & $\langle r^{-1} \rangle$ & $\rho_0^{-1}$\\
        \hline
        2 &$-0.060$[50.5]&12.88[39.2]&14.43[56.0]&208.3[96.7]&208.3[96.7]&0.113[31.1]&0.069[57.8]\\
        1 (DOS)&$-0.167$[37.6]&5.984[35.3]&5.196[43.8]&45.00[57.5]&27.00[74.5]&0.244[48.4]&0.192[17.1]\\
        1.502 (GS)&$-0.094$[22.8]&9.228[0.24]&9.258[0.08]&107.0[1.03]&85.71[19.1]&0.158[3.80]&0.108[34.3]\\
        \hline
    \end{tabular} 
    \caption{Same as Table~\ref{tab:coul1} for the state $|1; 2; 0^+\rangle$ with $E_\textrm{acc}=-0.121$, $\nu=1.92$ and $\sum_s d_s(2)^2 = 0.74$. \label{tab:coul2}}
\end{center}
\end{table}

\begin{table}[H]
\begin{center}
    \begin{tabular}{cccccccc}
        \hline
        $\phi$ & $E$ & $\langle r \rangle$ & $\rho_0$ &$\langle r^2 \rangle$ & $\rho_0^2$ & $\langle r^{-1} \rangle$ & $\rho_0^{-1}$\\
        \hline
        2 &$-0.060$[19.0]&13.39[8.27]&14.43[16.7]&208.3[12.1]&208.3[12.1]&0.095[12.0]&0.069[35.7]\\
        1 (DOS)&$-0.094$[26.6]&9.582[22.5]&9.238[25.3]&106.7[42.6]&85.33[54.1]&0.132[22.9]&0.108[0.49]\\
        1.502 (GS)&$-0.074$[0.05]&11.44[7.49]&11.70[5.38]&152.1[18.2]&137.0[26.3]&0.111[2.95]&0.085[20.7]\\
        \hline
    \end{tabular} 
    \caption{Same as Table~\ref{tab:coul1} for the state $|1; 2; 2^+\rangle$ with $E_\textrm{acc}=-0.074$, $\nu=1.22$ and $\sum_s d_s(2)^2 = 0.91$. \label{tab:coul3}}
\end{center}
\end{table}

\subsection{Gaussian three-body potential}
\label{sec:gauss} 

A repulsive Gaussian three-body potential can be added to attractive Gaussian two-body potentials to study clusters of helium atoms \cite{gatt11}. As we consider here only a three-body interaction, we use the following attractive version  
\begin{equation}\label{gauss1}
V_\textrm{G}\left( r_{\{123\}} \right) = -a\, e^{-b\,r^2_{\{123\}}}  .
\end{equation}
The relative spectra depend only on the ratio $a/b$ of the two dimensionless positive constants $a$ and $b$. By playing with these two parameters, the absolute spectra can be fixed at will. In theory, a physically relevant system must have a positive mass. This implies that the depth of the potential well must be much less than three times the mass of the particles (with our reduced units, this means $a \ll 3$). As we are only interested by tests about the quality of the spectra, we fix $a=200$ and $b=1$. This choice allows the existence of several bound states. Results for states with $Q^* \le 2$ are given in Tables~\ref{tab:gauss1} to \ref{tab:gauss3}. Comments about these results are:

\begin{itemize}
  \item Three bound states are found for $Q^* \le 2$ which correspond to the three first bound states computed numerically, with the good hierarchy.
  \item The original method ($\phi=2$) gives good results for the energies and the observables with reasonable relative errors. All energies computed with $\phi=2$ are upper bounds and the variational character is generally lost otherwise, as expected. 
  \item With $\phi=\phi_\textrm{GS}$, the energies are improved and the hierarchy respected but the observables are only very slightly modified. 
  \item The values of $\phi_\textrm{GS}$ and $\phi_\textrm{DOS}$ are close and give similar results. 
  \item The accuracy of observables $\langle r^k \rangle$ is generally better with formula~(\ref{obs2}) than with approximation~(\ref{fr}). But the relation $\rho_0^k$ can very easily give at least a rough approximation, and sometimes a good one.
  \item For the original method ($\phi=2$), $\nu\approx 1$ and $\sum_s d_s(Q^*)^2 \approx 1$. This shows a good quality for the ET wavefunction, in agreement with the quality of energies and observables.
\end{itemize} 

\begin{table}[H]
    \begin{center}
    \begin{tabular}{cccccccc}
        \hline
         $\phi$ & $E$ & $\langle r \rangle$ & $\rho_0$ &$\langle r^2 \rangle$ & $\rho_0^2$ & $\langle r^{-1} \rangle$ & $\rho_0^{-1}$\\
        \hline
        2 &$-103.2$[2.21]&0.292[2.48]&0.317[5.85]&0.101[5.63]&0.101[5.63]&4.354[1.77]&3.151[26.4]\\
        1.954 (DOS)&$-104.5$[0.91]&0.292[2.62]&0.315[4.89]&0.100[5.89]&0.099[7.33]&4.360[1.91]&3.179[25.7]\\
        1.922 (GS)&$-105.5$[0.0]&0.292[2.71]&0.313[4.22]&0.100[6.07]&0.098[8.51]&4.364[2.01]&3.200[25.2]\\
        \hline
    \end{tabular} 
    \caption{Results for potential~(\ref{gauss1}). Energy $E$ and three observables for the state $|1; 0; 0^+\rangle$ with $\phi=2$, $\phi_\textrm{DOS}$ and $\phi_\textrm{GS}$. Observables $\langle r^k \rangle$ are computed with (\ref{obs2}) and (\ref{fr}). Relative errors (in \%) are written between square brackets. The accurate value of the energy is $E_\textrm{acc}=-105.5$. For $\phi=2$, $\nu=0.95$ and $\sum_s d_s(0)^2 = 0.998$. \label{tab:gauss1}}
    \end{center}
\end{table}

\begin{table}[H]
\begin{center}
    \begin{tabular}{cccccccc}
        \hline
        $\phi$ & $E$ & $\langle r \rangle$ & $\rho_0$ &$\langle r^2 \rangle$ & $\rho_0^2$ & $\langle r^{-1} \rangle$ & $\rho_0^{-1}$\\
        \hline
        2 &$-47.33$[15.7]&0.392[7.44]&0.439[3.76]&0.193[13.9]&0.193[13.9]&3.725[7.16]&2.278[34.5]\\
        1.954 (DOS)&$-49.71$[11.4]&0.390[7.78]&0.433[2.43]&0.191[14.5]&0.188[16.1]&3.739[7.55]&2.308[33.6]\\
        1.922 (GS)&$-51.40$[8.44]&0.389[8.01]&0.429[1.50]&0.190[15.0]&0.184[17.6]&3.749[7.83]&2.329[33.0]\\
        \hline
    \end{tabular} 
    \caption{Same as Table~\ref{tab:gauss1} for the state $|1; 2; 0^+\rangle$ with $E_\textrm{acc}=-56.14$, $\nu=0.95$ and $\sum_s d_s(2)^2 = 0.96$. \label{tab:gauss2}}
\end{center}
\end{table}

\begin{table}[H]
\begin{center}
    \begin{tabular}{cccccccc}
        \hline
        $\phi$ & $E$ & $\langle r \rangle$ & $\rho_0$ &$\langle r^2 \rangle$ & $\rho_0^2$ & $\langle r^{-1} \rangle$ & $\rho_0^{-1}$\\
        \hline
        2 &$-47.33$[7.86]&0.407[3.24]&0.439[4.30]&0.193[7.35]&0.193[7.35]&3.116[2.30]&2.278[25.2]\\
        1.843 (DOS)&$-51.43$[0.13]&0.405[3.85]&0.429[2.00]&0.190[8.51]&0.184[11.4]&3.135[2.95]&2.329[23.5]]\\
        1.922 (GS)&$-49.36$[3.91]&0.406[3.54]&0.434[3.16]&0.192[7.93]&0.189[9.36]&3.126[2.62]&2.303[24.4]\\
        \hline
    \end{tabular} 
    \caption{Same as Table~\ref{tab:gauss1} for the state $|1; 2; 2^+\rangle$ with $E_\textrm{acc}=-51.36$, $\nu=0.94$ and $\sum_s d_s(2)^2 = 0.992$. \label{tab:gauss3}}
\end{center}
\end{table}

\section{Concluding remarks}
\label{sec:cr}

The envelope theory (ET) is a very simple approximation method to solve eigenvalue quantum equations for $N$ identical particles in $D$ dimensions for two-body interactions \cite{sema13a}. The method can be extended to treat potentials including a special type of many-body forces where the radial variables are the sum of squares of relative two-body distances \cite{sema18}, which are similar to the hyperradius variables introduced in the hyperspherical harmonic expansion method. The main interest of the method is that the computational cost is very low and independent from $N$. Moreover, analytical bounds of the energy can be obtained in the most favourable cases. Good results for eigenenergies and some observables can be obtained within the ET for various few-body systems, specially with the modified procedure \cite{sema15a,lore23,sema15b}. In this work, we check if the same situation happens for three simple nonrelativistic three-body Hamiltonians containing a three-body interaction of the type described above. 

For the two potentials without singularity, the linear and the Gaussian ones, the original version of the ET provides good results for the eigenvalues, the eigenfunctions and the considered observables. The accuracy obtained for these observables is better with the use of the formula~(\ref{obs2}) requiring the knowledge of the wavefunction than with the simpler approximation~(\ref{fr}). The introduction in the global quantum number of the parameter $\phi$ raising partly the non-physical degeneracy can improve the eigenvalues but generally not the observables. The price to pay is nevertheless to have the knowledge of one eigenvalue of this system, the ground state in this work, to fix the value of $\phi$. This is not a necessary easy task. Unfortunately, the prediction of $\phi$ using the dominantly orbital state (DOS) method which works very well for two-body potentials \cite{sema15b,lore23} can give poor results for three-body potentials. The most relevant information is knowing whether to decrease or increase $\phi$. For the Coulomb potential, the original version of the ET cannot provide good results for the eigenvalues. A dramatic improvement is possible with $\phi$ but a reasonable value for this parameter cannot be predicted by the coupling with the DOS method. 

As mentioned above, the value of $\phi$ can also be computed with the knowledge of the ground state of the system. In the case of a nonrelativistic kinematics, good approximations for the ground state of systems with identical bosons or fermions can be quite easily computed within the hyperspherical harmonic expansion method by taking into account only the hyperspherical harmonics of minimal order \cite{fabr79}. Such results must certainly be sufficient to compute a reasonable estimation of $\phi$. The situation is specially suited for the cases studied here: three-body systems with a potential depending only on the hyperradius. But there is no conceptual barrier to use the hyperspherical harmonic expansion in its lowest order in combination with the envelope theory for $N$-body systems with the type of $K$-forces considered in this paper. Such a work is certainly worth investigating.

We have remarked that both states $|1; 3; 1^-\rangle$ and $|1; 3; 3^-\rangle$ have the same energy for the three potentials but different values for observables. This degeneracy appears because Hamiltonians of types~(\ref{H123}) depend only on the hyperradius $r_{\{123\}}$. It disappears with the presence of two-body potentials. This will be discussed elsewhere \cite{chev23}.  

Only three simple Hamiltonians have been tested for three-body forces. One can imagine numerous supplementary tests with other potentials, semi-relativistic kinematics \cite{silv10}, the mixing of two-body forces, or the superposition of repulsive and attractive interactions \cite{lore23}. It could also be very interesting to test the quality of the method with a greater number of particles than three, as in \cite{sema15a}. The role of the dimensionality could be explored, in particular for $D=1$ \cite{sema19}. At last, the study of systems with different particles \cite{chev22} could be useful to extend the range of problems that can be investigated with the ET.  

Numerous methods to solve quantum many-body systems rely on the construct of good approximations for the wavefunctions. This allows generally systematic improvements of the eigenvalues and eigenvectors by adding in the trial states more and more parameters (for instance, in expanding the bases). The accuracy can then be controlled if the method is variational. The situation is very different with the ET. Even if the system considered allows a variational approximation, the only modification possible can improve the eigenvalues but cannot guarantee the preservation of their variational character nor the improvement of eigenfunctions and observables. Nevertheless, the method is so simple to use with a low computational cost independent from the number of particles, that is worth considering it. Indeed, the solutions obtained can have a very good accuracy with two- or three-body potentials in some situation and can then be very useful, for instance in the Large-$N$ limit of QCD where the number of particles considered can be arbitrary large \cite{buis12,buis22}. The results from ET can also be used as tests for numerical heavy calculations \cite{sema15a}, or even for pedagogical purposes \cite{sema23}.  

\begin{acknowledgments}
L.C. and C.C. would like to thank the Fonds de la Recherche Scientifique - FNRS for the financial support. C.T. would thank the University of Mons for her Initiation Research Grant. This work was also supported by the IISN under Grant Number 4.45.10.08. 
\end{acknowledgments} 

\appendix

\section{Symmetrisation of the wavefunctions}
\label{sec:sym}

In terms of the permutation operators $P_{ij}$ for the particles $i$ and $j$, the symmetriser for three-body systems is given by
\begin{equation}
\label{sym1}
S_\sigma = 1+\sigma\,P_{12}+\sigma\,P_{13}+\sigma\,P_{23}+P_{13}\,P_{12}+P_{23}\,P_{12},
\end{equation}
with $\sigma = 1$ for symmetric states and $\sigma = -1$ for antisymmetric states. With the choice of the Jacobi coordinates $\bm x_1$ and $\bm x_2$, we have $P_{12}\,\psi_s=(-1)^{l_1}\,\psi_s$, so
\begin{equation}
\label{sym2}
S_\sigma\,\psi_s = \left(\sigma+(-1)^{l_1}\right)\left(\sigma+P_{13}+P_{23}\right)\psi_s.
\end{equation}
The action of $P_{13}$ and $P_{23}$ can be computed with the Brody-Moshinsky coefficients $\langle nln'l';L|n_1 l_1 n_2 l_2;L\rangle_\beta$ described in \cite{silv85}. After some calculations, the correctly symmetrised states are written
\begin{align}
\label{sym3}
\psi &= A\,\left(\sigma+(-1)^{l_1}\right) \Big(\sigma\,\left[\varphi_{n_1 l_1}(\bm y_1)\,\varphi_{n_2 l_2}(\bm y_2)\right]^L + (-1)^{l_1+l_2-L}  \nonumber \\
& \times\, \sum_{n,l,n',l'}\left( \langle nln'l';L|n_1 l_1 n_2 l_2;L\rangle_{5\pi/3} + \langle nln'l';L|n_1 l_1 n_2 l_2;L\rangle_{\pi/3} \right) \left[\varphi_{n'l'}(\bm y_1)\,\varphi_{n l}(\bm y_2)\right]^L \Big).
\end{align}
The coefficient $A$ is used for the final normalisation. For instance, the normalised completely symmetric states with $Q^* = L_\textrm{max} \le 3$, a good total angular momentum $L$ and a parity $P=(-1)^{Q^*}$, written in the notation~(\ref{sym4}), are 
\begin{align}
\label{sym5}
|1;0;0^+\rangle &= |0,0,0,0;0\rangle, \nonumber \\
|1;2;0^+\rangle &= \frac{1}{\sqrt{2}} \left( |1,0,0,0;0\rangle + |0,0,1,0;0\rangle \right),\nonumber \\
|1;2;2^+\rangle &= \frac{1}{\sqrt{2}} \left( |0,2,0,0;2\rangle + |0,0,0,2;2\rangle \right),\nonumber \\
|1;3;1^-\rangle &= -\frac{1}{2} |0,0,1,1;1\rangle + \frac{1}{\sqrt{3}} |0,2,0,1;1\rangle + \sqrt{\frac{5}{12}} |1,0,0,1;1\rangle, \nonumber \\
|1;3;3^-\rangle &= \frac{1}{2} |0,0,0,3;3\rangle - \frac{\sqrt{3}}{2} |0,2,0,1;3\rangle. 
\end{align}
Symmetric state with $[Q^*=1]$, $[Q^*=2,L=1]$, $[Q^*=3,L\ \textrm{even}]$ are forbidden. It is easy to check the equality~(\ref{r2}) for the states~(\ref{sym5}) using (\ref{laj}) and (\ref{rk4}).

\section{Special integrals}
\label{sec:sint}

Let us note  
\begin{equation}
\label{rk1}
\langle r^k \rangle_{nl} = \int r^k \left|\varphi_{n l m}(\lambda\, \bm r)\right|^2 d\bm r,
\end{equation}
where $\varphi_{n l m}(\lambda\, \bm r)$ is a 3-dimensional harmonic oscillator wavefunction with the scale parameter $\lambda$. This integral is independent from the magnetic quantum number $m$. Using properties of the generalised Laguerre polynomials $L_m^\alpha(x)$ \cite{gradshteyn}, one gets 
\begin{equation}
\label{rk2}
\langle r^k \rangle_{nl} = \frac{1}{\lambda^k} \frac{\Gamma(n+l+\frac{3}{2})}{n!} \sum_{p,q=0}^n (-1)^{p+q}\, C_n^p\, C_n^q \frac{\Gamma(l+p+q+\frac{k+3}{2})}{\Gamma(l+p+\frac{3}{2})\,\Gamma(l+q+\frac{3}{2})}.
\end{equation}
Another form is given by  \cite{wolfram}
\begin{equation}
\label{rk3}
\langle r^k \rangle_{nl} = \frac{1}{\lambda^k} \frac{\Gamma\left( l+\frac{k+3}{2} \right)\Gamma\left( n-\frac{k}{2} \right)}{n!\,\Gamma\left( -\frac{k}{2} \right)\Gamma\left( l+\frac{3}{2} \right)}\, {_3F_2}\left(-n, l+\frac{k+3}{2}, \frac{k+2}{2}, \frac{k+2}{2}-n, l+\frac{3}{2} ; 1\right),
\end{equation}
where ${_3F_2}$ is a hypergeometric function. But this formula is not valid for $k$ positive and even. When $k=2$, using the virial theorem, it is easy to obtain
\begin{equation}
\label{rk4}
\langle r^2 \rangle_{nl} = \frac{2n+l+\frac{3}{2}}{\lambda^2}.
\end{equation}

\end{document}